\documentclass[twocolumn,english,aps,prb,floatfix,preprintnumbers,showpacs,amsfonts,amssymb,superscriptaddress]{revtex4}
\usepackage{ae,aecompl}
\usepackage[T1]{fontenc}
\usepackage[latin1]{inputenc}
\usepackage{longtable}
\usepackage{booktabs}
\usepackage{amsmath}
\usepackage{graphicx}
\usepackage{amssymb}

\makeatletter

\providecommand{\tabularnewline}{\\}

\usepackage{amscd}
\usepackage{bm}

\usepackage{babel}
\makeatother

\begin{document}

\title{Entanglement Entropy, Conformal Invariance, and the Critical Behavior
of the Anisotropic Spin-$S$ Heisenberg Chains: A DMRG study}

\author{J.~C.~Xavier}

\affiliation{Instituto de F\'{\i}sica, Universidade Federal de Uberlândia, Caixa
Postal 593, 38400-902 Uberl\^andia, MG, Brazil}

\date{\today{}}

\begin{abstract}
Using the density-matrix renormalization-group, we investigate the critical
behavior of the anisotropic Heisenberg chains with spins up to $S=9/2$. We show
that through the relations arising from the conformal invariance and the DMRG technique
it is possible to obtain accurate finite-size estimates of the conformal anomaly $c$, the sound velocity
$v_{s}$, the anomalous dimension $x_{bulk}$, and the surface exponent
$x_{s}$ of the anisotropic spin-$S$ Heisenberg chains with relatively good accuracy without
fitting parameters. Our results indicate  that the entanglement entropy $S(L,l_{A},S)$ of the spin-$S$
Heisenberg chains satisfies the relation $S(L,l_{A},S)-S(L,l_{A},S-1)=1/(2S+1)$ for
$S>3/2$ in the thermodynamic limit.
\end{abstract}

\pacs{03.67.Mn, 05.70.Jk, 75.10.Jm }

\maketitle

\section{INTRODUCTION}

Conformal invariance plays an important role in the study of critical
one-dimensional quantum systems. Several constraints appear if we
assume that the critical systems are conformal invariant. \citealp{critical1,critical2,critical3}
In particular, the possible classes of critical behavior of the one-dimensional
quantum systems are indexed by the conformal anomaly (or central charge)
$c$ as well as the values the anomalous dimensions $x_{bulk}^{\alpha}$
of the primary scaling operators $O^{\alpha}$.\citealp{critical1,critical2,critical3,anomaly2,anomaly1}

In the mid eighties of last century, it was shown that
the conformal anomaly $c$ can be extracted from the large-$L$ behavior
of the ground state energy $E_{0}(L)$. The ground state energy of
a system of size $L$ behaves as\citealp{anomaly1,anomaly2}

\begin{equation}
\frac{E_{0}}{L}=e_{\infty}+\frac{f_{\infty}}{L}-\frac{v_{s}\pi c}{\delta6L^{2}},\label{eq:1}\end{equation}
where the constant $\delta=1$(4) for the systems under periodic (open/fixed)
boundary conditions, $v_{s}$ is the sound velocity, $e_{\infty}$
is the bulk ground state energy per site, and $f_{\infty}$ is the
surface free energy, which vanishes for the systems under periodic
boundary conditions (PBC). The structure of the higher energy states,
for a system with periodic (open) boundary conditions, are related
with the anomalous dimensions $x_{bulk}^{\alpha}$ (surface exponents
$x_{s}^{\alpha}$). There are a tower of states in the spectrum of
the Hamiltonian with energies $E_{m,m'}^{\alpha}(L)$ given by\citealp{dimensions,surfexp}

\begin{equation}
E_{m,m'}^{\alpha}(L)-E_{0}(L)=\frac{2\pi v_{s}}{\eta L}(x+m+m'),\label{eq:2}\end{equation}
where $m,m'=0,1,2,...$, the constant $\eta=1$(2) and $x=x_{bulk}^{\alpha}(x_{s}^{\alpha})$
for the systems with periodic (open) boundary conditions.

The critical behavior of several models were studied using the conformal
invariance relations above. In particular, the spin-1/2 $XXZ$ 
chain is exactly soluble. For this reason, in this case, it is possible to obtain the spectrum of
the energies of very large system sizes (solving the Bethe ansatz
equations). Due to this fact, very accurate estimates of $c$, $v_{s}$,
$x_{bulk}^{\alpha}$ and $x_{s}^{\alpha}$ were obtained for the spin-1/2
$XXZ$  chain. \citealp{chico1,chico2} However, very few
models are exactly soluble. And, in general, it is possible obtain
the eigenspectrum only by exact numerical diagonalizations.

The Lanczos method of exact diagonalization (ED), can be used to extract
the ground-state energies of the Halmiltonians. However, with ED it
is possible to consider small system sizes, since the Hilbert space
grows exponentially with the system size. Even though, relative good
estimates of conformal anomaly $c$ and the anomalous dimensions $x_{bulk}^{\alpha}$
can be obtained with ED (see, for example, Refs. \onlinecite{spinschico,Rittenberg}).

The density-matrix renormalization-group (DMRG)\citealp{white} is
a powerful numerical technique that can be used to study very large
one-dimensional systems. In principle, we can use the DMRG technique
and the conformal invariance relations to obtain high accuracy estimates
of $c$, $x_{bulk}^{\alpha}$ and $x_{s}^{\alpha}$. Since with the
DMRG it is possible to obtain the ground states energies of large system
sizes in a controlled way. However, it is not easy to estimate the sound
velocity $v_{s}$ with the DMRG technique, differently of the ED (it
is possible to estimate $v_{s}$ with the ED by the behavior of the
energies in\emph{ different momentum sectors}). This is the major
reason why the conformal invariance relations above were not explored
in the studies of critical systems by the DMRG technique. \cite{comm2}

In recent years, this has changed because Calabrese and Cardy  related the
entanglement entropy of one-dimensional critical systems with the conformal anomaly
$c$\citealp{cardyentan} (see also Refs. \onlinecite{cold,affleckboundary,cvidal}). 
As we will see, due to this new relation  between the confomal anomaly and the
entanglement entropy, it is now possible to obtain the 
\emph{finite-size estimates of} $c$, $v_{s}$, $x_{bulk}$, and $x_{s}$
in a systematic way with the DMRG technique. 
The entanglement entropy can be defined as following. \cite{Schumacher} Consider a
one-dimensional system with size $L$ and composed by two subsystems
$A$ and $B$ of sizes $l_{A}$ and $l_{B}=L-l_{A},$ respectively.
The entanglement entropy is defined as the von Neumann entropy $S(L,l_{A})=-Tr\rho_{A}\ln(\rho_{A})$,
associated with the reduced density matrix $\rho_{A}$. For the critical
one-dimensional systems the entanglement entropy behaves as\citealp{cardyentan} 

\begin{equation}
S(L,l_{A})=\frac{c}{3\eta}\ln\left(\frac{\eta L}{\pi}\sin\left(\frac{\pi l_{A}}{L}\right)\right)+c_{1}-(1-\eta)s_{b},\label{eq:3}\end{equation}
where $s_{b}$ is the boundary entropy,\citealp{affleckboundary}
$c_{1}$ is a non-universal constant and $\eta=1$(2) for the systems
under periodic (open/fixed) boundary conditions. 

Note that we can estimate the conformal anomaly $c$ using 
Eq. \ref{eq:3} \emph{without the knowledge} of the sound velocity $v_{s}$. Indeed,
some authors have plotted $S(L,l_{A})$ as function of
$\ln\left(\frac{L}{\pi}\sin\left(\frac{\pi l_{A}}{L}\right)\right)$
in order to estimate $c$ by \emph{a numerical fit}.
Here, we use basically the same route that 
Zhao {\it et al.} used in Ref. \onlinecite{peschel}   to extract $c$
in an investigation of the XXZ Heisenberg chains with defects.
We will estimate $c$ by measuring the entanglement entropy of two systems 
with different sizes. With this procedure, we can obtain a 
\emph{finite-size} estimate of $c$ without a numerical fit. 
We  have observed that this procedure  gives errors smaller
than the ones obtained by the fitting procedure (in same cases,  the error
is one order of magnitude smaller).

It is important to point out that  some authors have used the DMRG 
to  obtain the { \it critical exponents} of several models \cite{dmrgcritical} 
through the asymptotic behavior of static correlation functions. 
This  numerical procedure, use more CPU time and, usually,  provides only rough 
estimates of the critical exponents. Thus, we see that 
a  procedure  to obtain accurate estimates  of the critical 
exponents (based in the energies behavior) with the DMRG is also highly desirable.

The main aim of this article is to show that 
if we explore all the \emph{three conformal 
invariance relations} above (Eqs. (1)-(3)) with the DMRG technique, then
it is  possible to obtain accurate \emph{finite-size estimates of} $c$, $v_{s}$, $x_{bulk}$, and $x_{s}$ 
in a systematic way with the DMRG technique  \emph{without any fitting
  parameter}. 
Several  authors have explored some of these relations with the DMRG.
However, they did not make use of all the three relations (with the exception 
of Ref. \onlinecite{peterentropy}).
We  would like to point out here that with the  multiscale entanglement
renormalization ansatz (MERA) it is also possible to extract these quantities
with a good accuracy\citealp{vidalising} (see also
Ref. \onlinecite{lucasetal}).

In order to illustrate the procedure that we use to extract $c$,
$v_{s}$, $x_{bulk}^{\alpha}$ and $x_{s}^{\alpha}$, we consider
the anisotropic spin-$S$ Heisenberg chains defined as 

\begin{equation}
H=\sum_{j}\left(s_{j}^{x}s_{j+1}^{x}+s_{j}^{y}s_{j+1}^{y}+\Delta s_{j}^{z}s_{j+1}^{z}\right),\label{eq:4}\end{equation}
were $\Delta=\cos(\gamma)$ is the anisotropy. It is well know that
this model at the isotropic point $\Delta=1$ is gapless (gapful)
for half-integer (integer) spins. \citealp{dagottosc1} For this
reason, we will consider only the half-integer spin cases. Although
even the models with integer spins become gapless at some critical anisotropy.
\citealp{spinschico}

We investigate the model defined above using the DMRG\citealp{white} method under
open boundary conditions (OBC) and PBC, keeping up to $m$=4000  states
per block in the final sweep. We have done $\sim6-9$ sweeps, and
the discarded weight was typically $10^{-9}-10^{-12}$ at that final
sweep. The dimension of the superblock in the last sweep can reach
up to 80 millions. In our DMRG procedure the center blocks 
are composed of (2$S$+1) states.

\section{RESULTS}

\subsection{S=1/2}

We will focus, firstly, in the anisotropic spin-1/2 Heisenberg chain,
which is exactly soluble (for a review see Ref. \onlinecite{chico1}).
For this case, we be able to compare our numerical estimates of $c$,
$v_{s}$, $x_{bulk}^{\alpha}$ and $x_{s}^{\alpha}$ with the exact
results. In order to avoid the logarithmic corrections (due to the
marginally irrelevant operator) which makes the finite-size analysis
more complicated, we decided to investigate the model at the
anisotropic point $\gamma=\pi/3$ (which corresponds to $\Delta=1/2$). 
Few results for other values of $\gamma$ were also explored in this 
work.

Before presenting the numerical results for the spin-1/2 chain, let us
describe some known results based on analytical approaches.\citealp{chico1,spinsaffleck,xxzhamer,xxxWoynarovich,peschel2}
The anisotropic spin-1/2 chain is critical for $-1<\Delta\le1$ with
central charge $c=1$, the anomalous dimension (surface exponent)
associated to the lowest eigenenergy in the sector with total spin  z-component
$S_z=1$ is given by $x_{bulk}=\frac{\pi-\gamma}{2\pi}$ ($x_{s}=\frac{\pi-\gamma}{\pi}$)
and the sound velocity is $v_{s}=\frac{\pi\sin(\gamma)}{2\gamma}$.
Very interesting  to note that for $\gamma=\pi/3$ the anisotropic spin-1/2
Heisenberg chain present a very peculiar ground state. For this
reason, some analytical expression for the two-point correlation
functions\cite{pair} and for the reduced density matrix were obtained.\cite{matrix}
Some exact results for the reduced density matrix with arbitrary $\gamma$ were also obtained by 
Alba {\it et al.} in Ref. \onlinecite{matrix2}.

\begin{figure}
\begin{centering}
\includegraphics[scale=0.3]{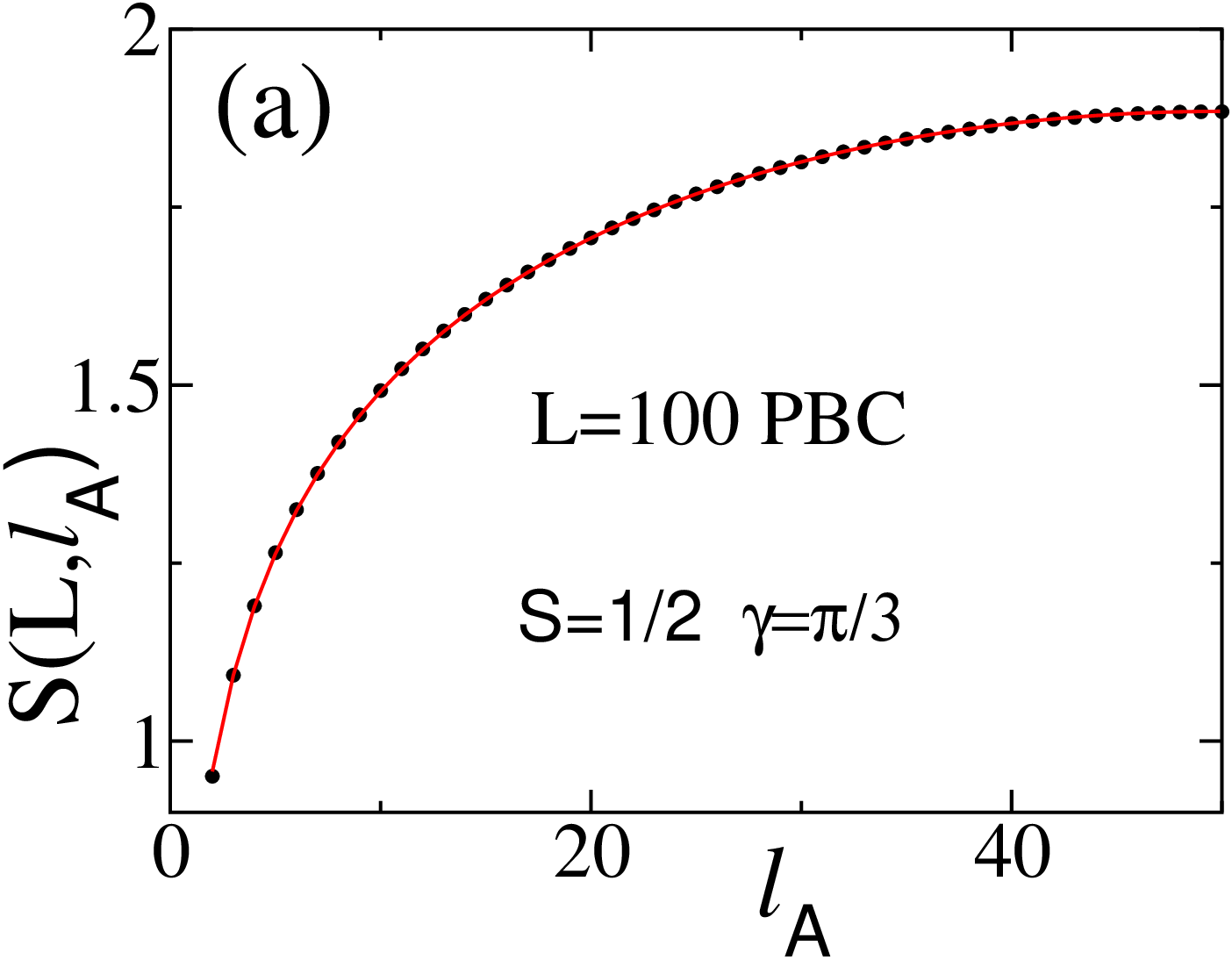}
\par\end{centering}

\begin{centering}
\includegraphics[scale=0.3]{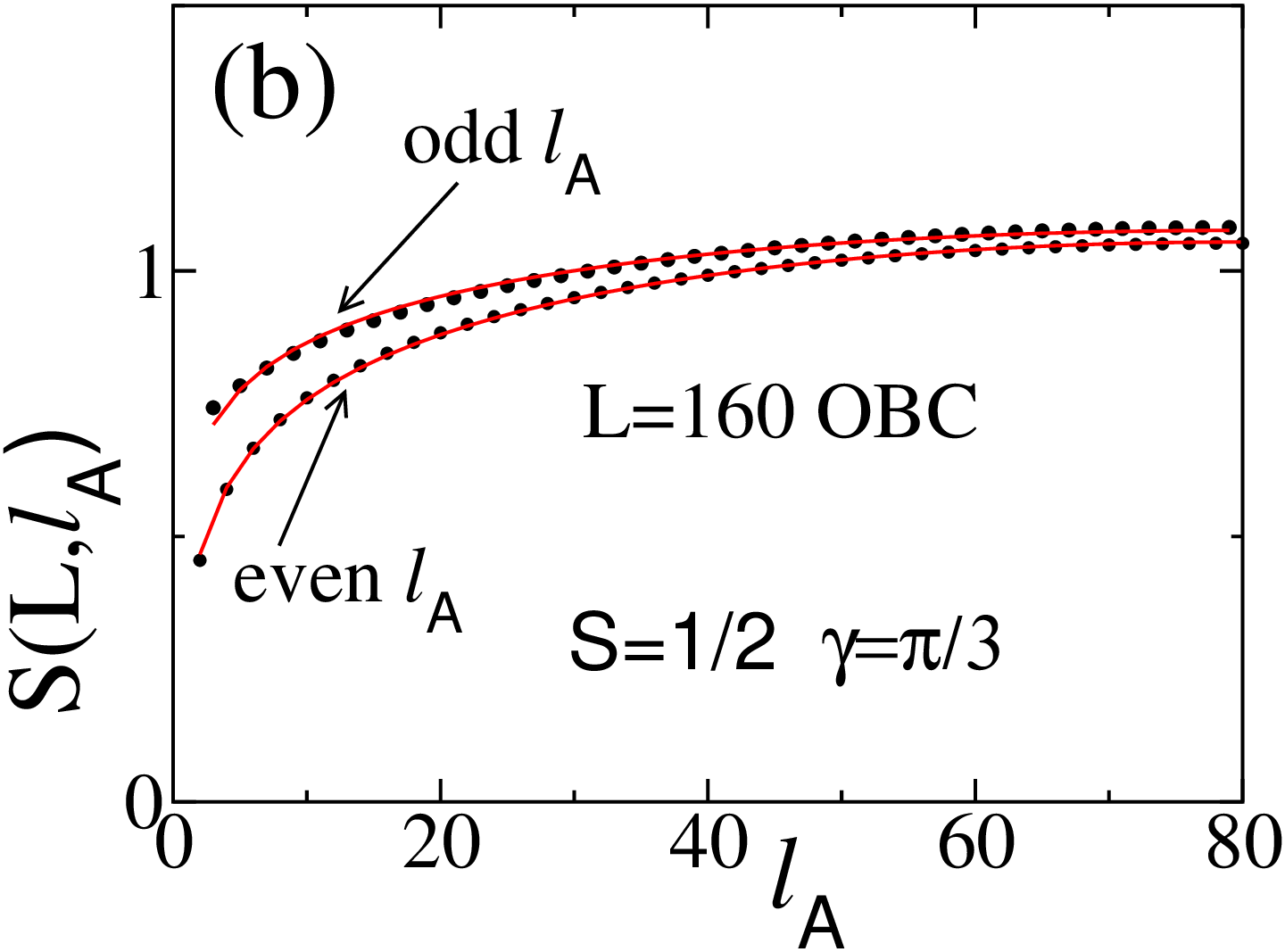}
\par\end{centering}

\caption{\label{fig1} (Color online). The entanglement entropy $S(L,l_{A})$
vs. $l_{A}$ for the anisotropic spin-1/2 Heisenberg with $\gamma=\pi/3$.
The circles are  numerical data and the red lines are fits to our
data using Eq. \ref{eq:3}  (see text). (a) $S(L,l_{A})$ vs. $l_{A}$
for a chain of size $L=100$ and PBC. (b) $S(L,l_{A})$ vs. $l_{A}$
for a chain of size $L=160$ and OBC. }

\end{figure}

In Fig. 1(a), we present the entanglement entropy $S(L,l_{A})$ as
function of the length $l_{A}$ for the anisotropic spin-1/2 Heisenberg
chain with PBC for a system of size $L=100$ and $\gamma=\pi/3$.\citealp{comm}
The dashed line in this
figure is a fit to our data using Eq. \ref{eq:3}. We show $S(L,l_{A})$
only for $\ell_{A}\le L/2$ since $S(L,L-l_{A})=S(L,l_{A})$. The
conformal anomaly obtained through this fit is $c=1.005$, which is
very close to the exact value $c_{exact}=1$. Estimates of the conformal
anomaly can also be acquired if we consider the system with OBC instead of PBC. 
In Fig. 1(b), we show $S(L,l_{A})$ as function of $l_{A}$ for the model
with OBC, $\gamma=\pi/3$ and $L=160$. Observe that for the OBC case,
$S(L,l_{A})$ exhibits even-odd oscillations as function of $l_{A}$.
In this case, for $l_{A}$ even (odd) the fit to our data using Eq.
\ref{eq:3} is $c=1.09$ ($c=0.77)$. These even-odd oscillations
have been reported in several works, \citealp{entropyaffleckosc,entroscs,roux,entroschollwock}
and it is expected to decay away from the boundary with a power law.
\citealp{entropyaffleckosc} Very interesting to note that even for
systems under PBC, the Rényi entropy also shows even-odd oscillations,
which have been proposed to have an universal scaling law. \citealp{entropyosc} 

The above route have been used by some authors to estimate the conformal
anomaly of some critical models such as the transverse Ising chain,
\citealp{entroschollwock} the  isotropic/anisotropic  spin-1/2 Heisenberg
chain,\citealp{entroschollwock,peterentropy} the SU(N)
chains,\citealp{peterentropy}
and as well as  a spin-3/2 fermionic cold atoms
with attractive interactions.\cite{roux} In order to obtain good estimates of $c$ with this
procedure we  need to consider large system sizes, mainly for the OBC case,
which  can demand  large computational effort in general.
Here, we use a slightly different route to obtain the finite-size estimates
of the conformal anomaly $c$, as mentioned before. In particular, for the OBC case, we will see
that with an extrapolation of the finite-size data, we are able to obtain estimates of
$c$ and  $x_{s}$  with good accuracy. Even considering relatively small
system sizes, which requires a reasonably small computational effort.

A simply way to extract the conformal anomaly from Eq. \ref{eq:3},
without any fitting parameter, is considering two systems with sizes $L$
and $L'$. Let us assume that these two systems are composed of two
subsystems of sizes $l_{A}=L/2$ and $l_{A}^{'}=L'/2$, respectively.
Thus, from Eq. \ref{eq:3}, we see that we can estimate $c$ by

\begin{equation}
c(L,L')=3\eta
\frac{S(L,L/2)-S(L',L'/2)}{\ln\left(L/L'\right)}
=\frac{3\eta\Delta S}{\ln\left(L/L'\right)}.
\label{eq:5}\end{equation}

\begin{table}
\caption{Finite-size estimates of the conformal anomaly $c^{PBC}$, the sound
velocity $v_{s}$, and the anomalous dimension $x_{bulk}$ for the
spin-1/2 anisotropic Heisenberg chain with PBC and $\gamma=\pi/3$.
We use $L'=L+16$ in Eq. \ref{eq:5}.}
\vspace*{0.2cm} 
\begin{tabular}{cccccc}

\toprule 
\midrule
$L$ & $c^{PBC}$ &  & $v_{s}$ & $x_{bulk}$ & \tabularnewline
\midrule 
\vspace*{0.05cm} 
16 & 0.99596 &  & 1.30675 & 0.33314 & \tabularnewline
\vspace*{0.05cm} 
32 & 0.99879 &  & 1.30103 & 0.33324 & \tabularnewline
\vspace*{0.05cm} 
48 & 0.99942 &  & 1.29994 & 0.33328 & \tabularnewline
\vspace*{0.05cm} 
64 & 0.99966 &  & 1.29955 & 0.33330 & \tabularnewline
\vspace*{0.05cm} 
80 & 0.99977 &  & 1.29937 & 0.33331 & \tabularnewline
 \vspace*{0.05cm} 
96 &  &  & 1.29927 & 0.33332 & \tabularnewline
\midrule 
exact & 1 &  & 1.29903 & 0.33333 & \tabularnewline
\midrule
\bottomrule
\end{tabular}
\end{table}

Note that previously works\cite{peschel,ren} also use the increment of the entropy, $\Delta S$,
to extract the conformal anomaly of the spin-1/2 XXZ Heisenberg model with defects/impurities.
L\"auchli and Kollath\cite{andreas} also  used the increment of entropy in
order to locate the quantum critical point  of the Bose-Hubbard chain. 

In Table I, we show the finite-size estimates of the conformal anomaly,
obtained by Eq. \ref{eq:5}, for the  anisotropic spin-1/2 Heisenberg
chain with PBC and $\gamma=\pi/3$. Similar results were also found
for $\gamma=\pi/6$ and $\gamma=\pi/8$. For comparison purpose, we also
present in this table the exact values of the conformal anomaly $c$,
the sound velocity $v_{s}$, and the anomalous dimension $x_{bulk}$.
Note that by using relatively small system sizes, we are able to estimate
$c$ with a small error ($\sim$2$\times10^{-4})$ \emph{without any
fitting parameter}.

Once the conformal anomaly $c$ is obtained, we can use Eq. \ref{eq:1}
to extract the sound velocity $v_{s}$. In particular, for the PBC
case the finite-size estimate of the sound velocity is obtained by\cite{comm3}

\begin{equation}
v_{s}(L)=\frac{6L}{\pi c}\left(Le_{\infty}-E_{0}(L)\right).\label{eq:6}\end{equation}
Note that in this equation, we also need the bulk ground-state energy
per site $e_{\infty}$. However, this is not a problem since with
the DMRG technique it is possible to obtain $e_{\infty}$ with a high
accuracy. 

The finite-size estimates of $v_{s}$ obtained  using
Eq. \ref{eq:6} are also presented in Table I. As observed in this
table, accurate results are also acquired for $v_{s}$. As we already
mentioned in the introduction, until very recently, it was not possible
to extract the sound velocity $v_{s}$, based only  in the large-$L$
behavior of the ground state energy with the DMRG technique. But, as we have observed,
with the use of the Eqs. \ref{eq:1} and \ref{eq:3} we can estimate
$v_{s}$, with good accuracy, considering relatively small system
sizes. 

Finally, using Eq. \ref{eq:2}, we extract the finite-size estimates
of the anomalous dimension/surface exponent by 

\begin{equation}
x(L)=\frac{\eta L\left(E_{1}(L)-E_{0}(L)\right)}{2\pi v_{s}},\label{eq:7}\end{equation}
where $E_{1}(L)$ is the ground state energy in the sector with total spin  z-component
$S_z=1$. In Table 1, we also show the finite-size estimates of
$x_{bulk}$ for the anisotropic spin-1/2 Heisenberg chain under PBC
and $\gamma=\pi/3$. Note again that accurate results are obtained. 

From the above results, we learn that it is possible to obtain good
estimates of $c$, $v_{s}$ and $x_{bulk}$, without any fitting parameters
(or extrapolations), considering relatively small system sizes under
PBC. On the other hand,  for the OBC case  is necessary to extrapolate the
finite-size estimates in order to obtain accurate results, as shown
below.

\begin{table}

\caption{Extrapolated and finite-size estimates of the conformal anomaly $c^{OBC}$
and the surface exponent $x_{s}$ for the anisotropic  spin-1/2 Heisenberg
chain with OBC and $\gamma=\pi/3$. We use $L'=L+20$ in Eq. \ref{eq:5}.
The extrapolated values were obtained by a numerical fit (see text).}
\vspace*{0.2cm} 
\begin{tabular}{cccccc}
\toprule
 \midrule
$L$ & $c^{OBC}$ & $x_{s}$ &  &  & \tabularnewline
\midrule 
60 & 1.0859 & 0.64298 &  &  & \tabularnewline
\vspace*{0.05cm}
80 & 1.0727 & 0.64809 &  &  & \tabularnewline
\vspace*{0.05cm}
100 & 1.0633 & 0.65131 &  &  & \tabularnewline
\vspace*{0.05cm}
120 & 1.0568 & 0.65353 &  &  & \tabularnewline
\vspace*{0.05cm}
140 & 1.0518 & 0.65517  &  &  & \tabularnewline
\vspace*{0.05cm}
160 & 1.0470 & 0.65642 &  &  & \tabularnewline
\midrule 
$\infty$ & 1.004 & 0.66658 &  &  & \tabularnewline
\midrule 
exact & 1 & 0.66666 &  &  & \tabularnewline
\midrule
\bottomrule
\end{tabular}
\end{table}

Let us now estimate the conformal anomaly considering the model
with OBC (Eq. \ref{eq:5} with $\eta=2$). In Table II, we present
the finite-size estimates of the conformal anomaly $c$ for the anisotropic
spin-1/2 Heisenberg chain with OBC and $\gamma=\pi/3$. As we observed
in this table, the finite-size estimates of $c^{OBC}$ are less accurate
than the ones found for the model with PBC, even  considering
larger systems for the OBC case. We can improve the estimate of $c^{OBC}$ extrapolating
the finite-size data to the infinite lattice value. We assume that
conformal anomaly behaves as

\begin{equation}
c^{OBC}(L)=c+a/L+b/L^{2}.\label{eq:8}\end{equation}
The extrapolated values in Table II, were obtained from a fit to our
data with this equation (we use only the last four points of our data
in the numerical fit). Note that the extrapolated estimate is quite
close to the exact value. As we see in this table, the estimate of
$c^{OBC}$ obtained with this procedure gives an absolute error of
about $10^{-3}$, which is quite good.

In a similar way, we also use Eq. \ref{eq:7} to obtain the surface
exponent $x_{s}$. In Table II, we also present the finite-size estimates
of $x_{s}$ for the anisotropic spin-1/2 Heisenberg chain with OBC
and $\gamma=\pi/3$. The extrapolated value of surface exponent 
$x_{s}$ was obtained by a numerical fit, as we did for the conformal anomaly. 
We assume that the surface exponent behaves as

\begin{equation}
x_{s}(L)=x_{s}+a/L+b/L^{2}.\label{eq:9}\end{equation}
As observed in the table, the extrapolated value of $x_{s}$ is also
very close to the exact one.

\subsection{S>1/2}

The results obtained above for the spin-1/2 chain are all known, and
we studied the spin-1/2 case  for the purpose of comparison/benchmark.
However, that study was highly important, because it established 
that the procedure used by us provides accurate results.
Now, let us consider the cases where $S>1/2$.

Before presenting the estimates of $c$, $x_{bulk}$, and $x_{s}$. We will
present an interesting behavior of the entanglement entropy $S(L,l_{A},S)$ 
for the spin-$S$ Heisenberg chains with spin $S>1/2$. We observed what appears to be
an universal  behavior of $S(L,l_{A},S)$ for $S>1/2$. 
Our numerical data indicate that the entanglement entropy of two systems with spins $S$ and
$S-1$ respectively, with $S>3/2$ are related by following the relation
\begin{equation}
S(L,l_{A},S)-S(L,l_{A},S-1)=\frac{1}{(2S-1)}+a_1,
\label{eq:10}
\end{equation}
where $a_1$ is a very small term. Note that this is equivalent to say
that the non-universal constant $c_1(S)$ (see Eq. \ref{eq:3})
satisfies the relation $c_1(S)-c_1(S-1)=\frac{1}{(2S-1)}+a_1$.

We have observed that constant $a_1$
decrease with the size of the systems. We were  not able to proof
that   $a_1$ is zero in the thermodynamic limit, although our  numerical 
results suggest that.  In particular, for a system with PBC (OBC), $\gamma=\pi/3$, and size $L=80$ (L=160) we found 
that $a_1\sim 10^{-3}$($a_1\sim 10^{-2}$). 
In the discussion below we neglect this term. 
Due to  the Eq. \ref{eq:10}, we can express the entanglement entropy 
$S(L,l_{A},S)$ of a systems with spin $S>3/2$, in terms of $S(L,l_{A},3/2)$ by the equation 
\begin{equation}
S(L,l_{A},S)=S(L,l_{A},3/2)+\sum_{j=0}^{(S-5/2)}\frac{1}{2(S-j)-1}.
\label{eq:11}
\end{equation}
In order to check the valid of the  Eq. \ref{eq:11}, we
investigate the following function 
\begin{equation}
W(L,l_{A},S)=S_{DMRG}(L,l_{A},S)-\sum_{j=0}^{(S-5/2)}\frac{1}{2(S-j)-1},
\label{eq:12}
\end{equation}
and we set $W(L,l_{A},3/2)=S_{DMRG}(L,l_{A},3/2)$.

\begin{figure}
\begin{centering}
\includegraphics[scale=0.3]{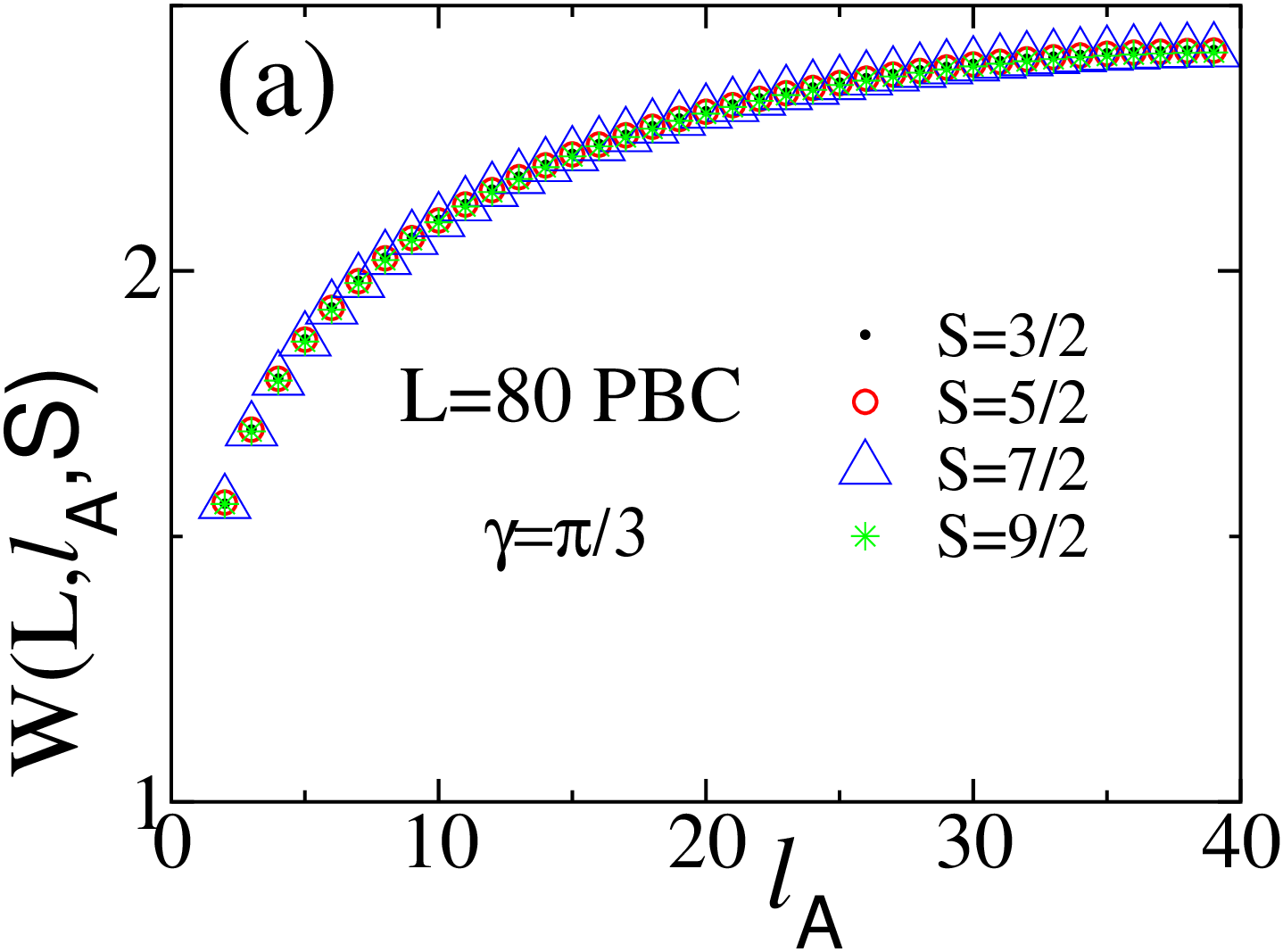}
\par\end{centering}

\begin{centering}
\includegraphics[scale=0.3]{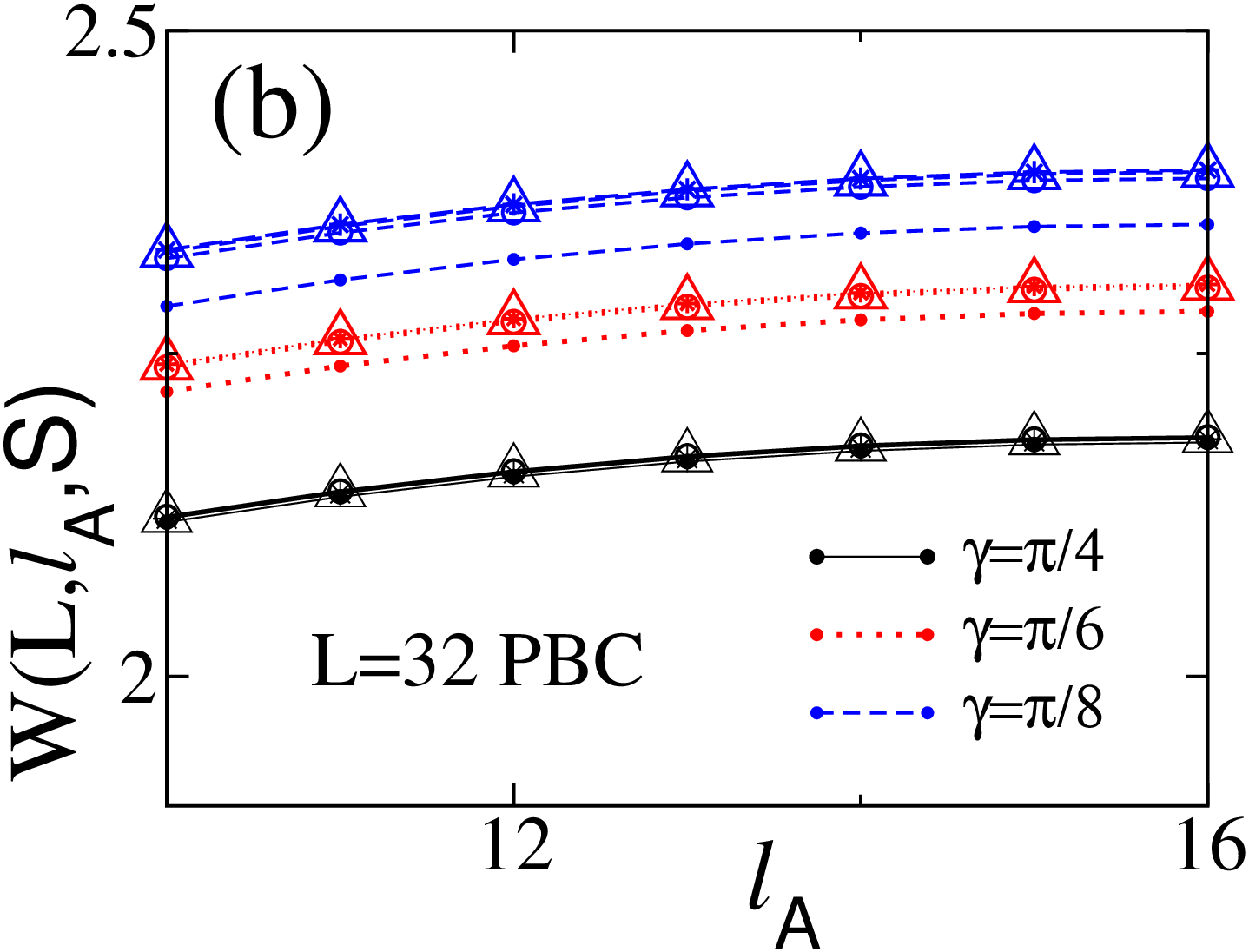}
\par\end{centering}

\begin{centering}
\includegraphics[scale=0.3]{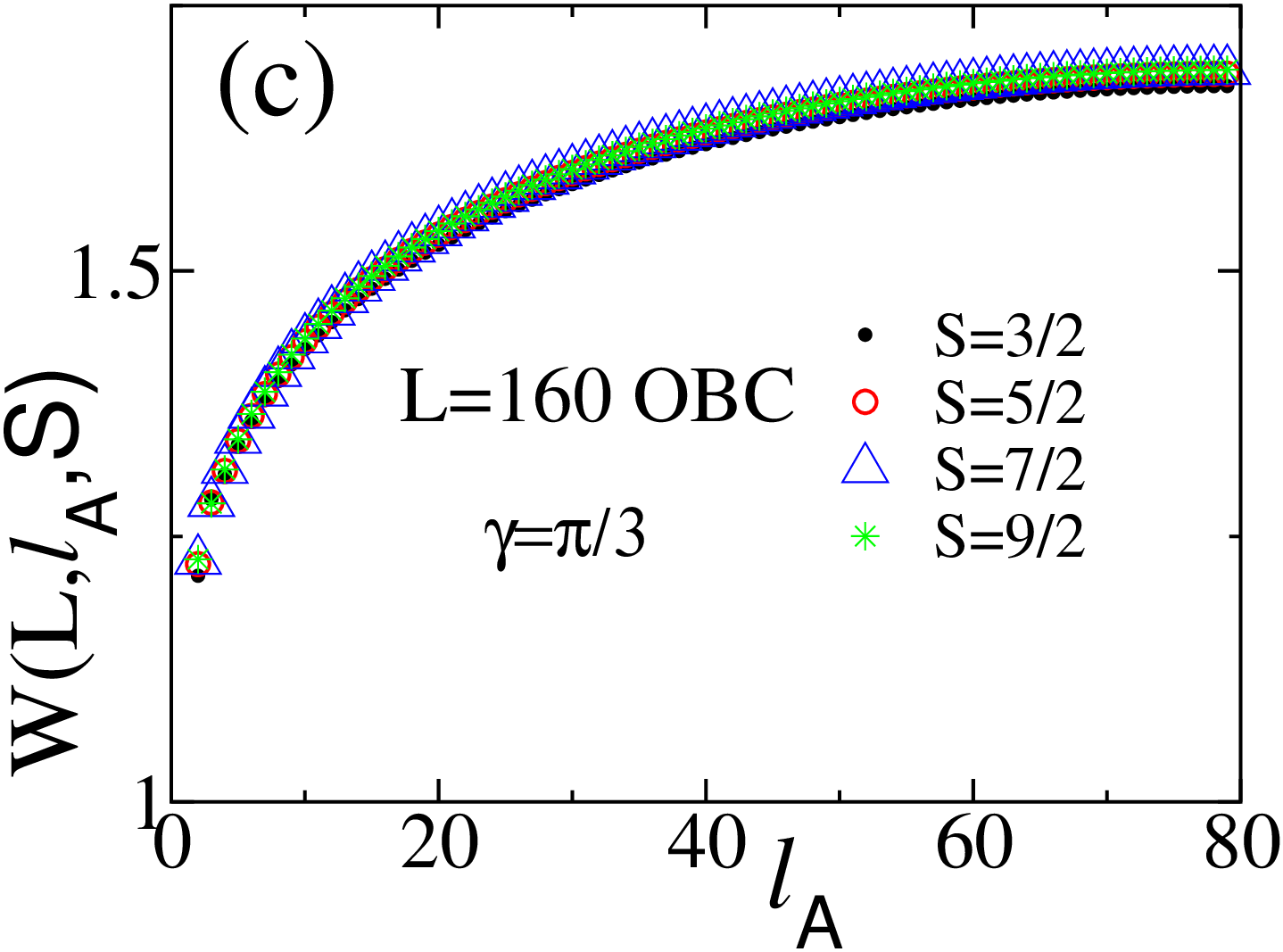}
\par\end{centering}

\caption{\label{fig2} (Color online).  The rescaled entanglement entropy $W(L,l_{A},S)$ 
vs. $l_{A}$ for the anisotropic spin-S Heisenberg for some
values of $S$ (see legend).
 (a) $W(L,l_{A},S)$ vs. $l_{A}$
for chains of size $L=80$, $\gamma=\pi/3$ and PBC.
 (b) $W(L,l_{A},S)$ vs. $l_{A}$ for chains of size $L=32$, PBC and some values
of $\gamma$ (see legend). The simbols   refer to the values of the spins (we
used the same simbols of figure (a)).
(c) $W(L,l_{A},S)$ vs. $l_{A}$
for  chains of size $L=160$ and OBC.
}

\end{figure}

Let us focus, firstly, in the PBC case. In Fig. \ref{fig2}(a), we
present the rescaled entanglement entropy  $W(L,l_{A},S)$ as function of 
$l_{A}$ for some values of spins  for systems with PBC, $L=80$ and
$\gamma=\pi/3$.  As observed in this figure, 
all curves   collapse onto a single universal scaling curve.
Actually, there is a small difference around $~10^{-3}$ between these
curves (see Fig.\ref{fig2}(b)). 
Similar results are also observed for other values of $\gamma$ for systems
of size $L=32$, as shown in Fig. \ref{fig2}(b). Note that as
$\gamma$ decrease the differences between the curves increase.
This is expected since  finite-size effects become stronger
 as $\gamma\rightarrow 0$ (or $\Delta\rightarrow 1$).
These results strongly indicate that Eq. \ref{eq:11} is valid for any 
values of spin $S>3/2$ and $\gamma$. Our numerical data
(not shown) support  also that
$S_{DMRG}(L,l_{A},3/2)=S_{DMRG}(L,l_{A},1/2)+11/18$ for the PBC case.
We will see below that if we consider the systems with OBC, 
the Eq. \ref{eq:11} still holds true for $S>3/2$.

\begin{figure}

\begin{centering}
\includegraphics[scale=0.3]{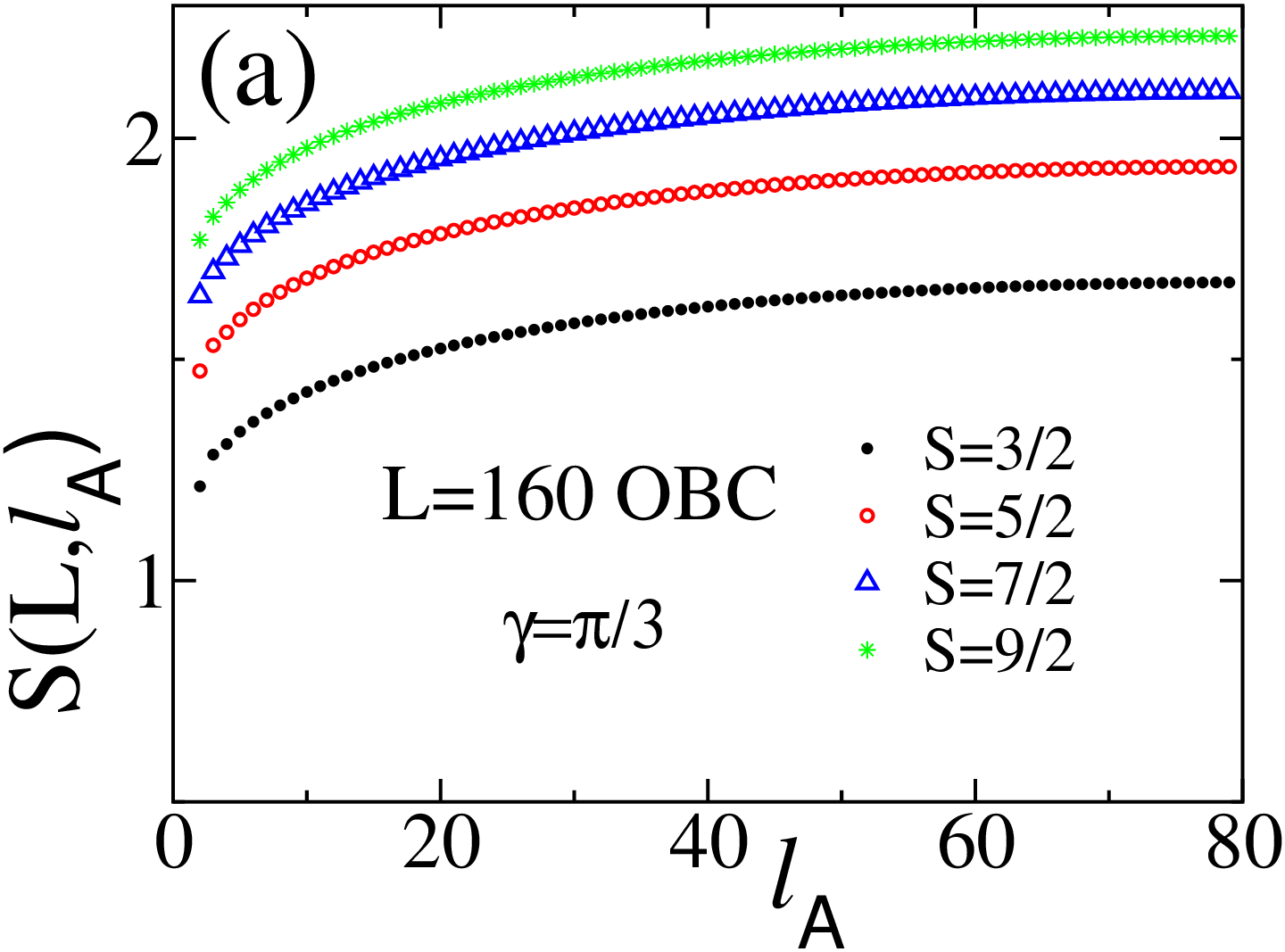}
\par\end{centering}

\begin{centering}
\includegraphics[scale=0.3]{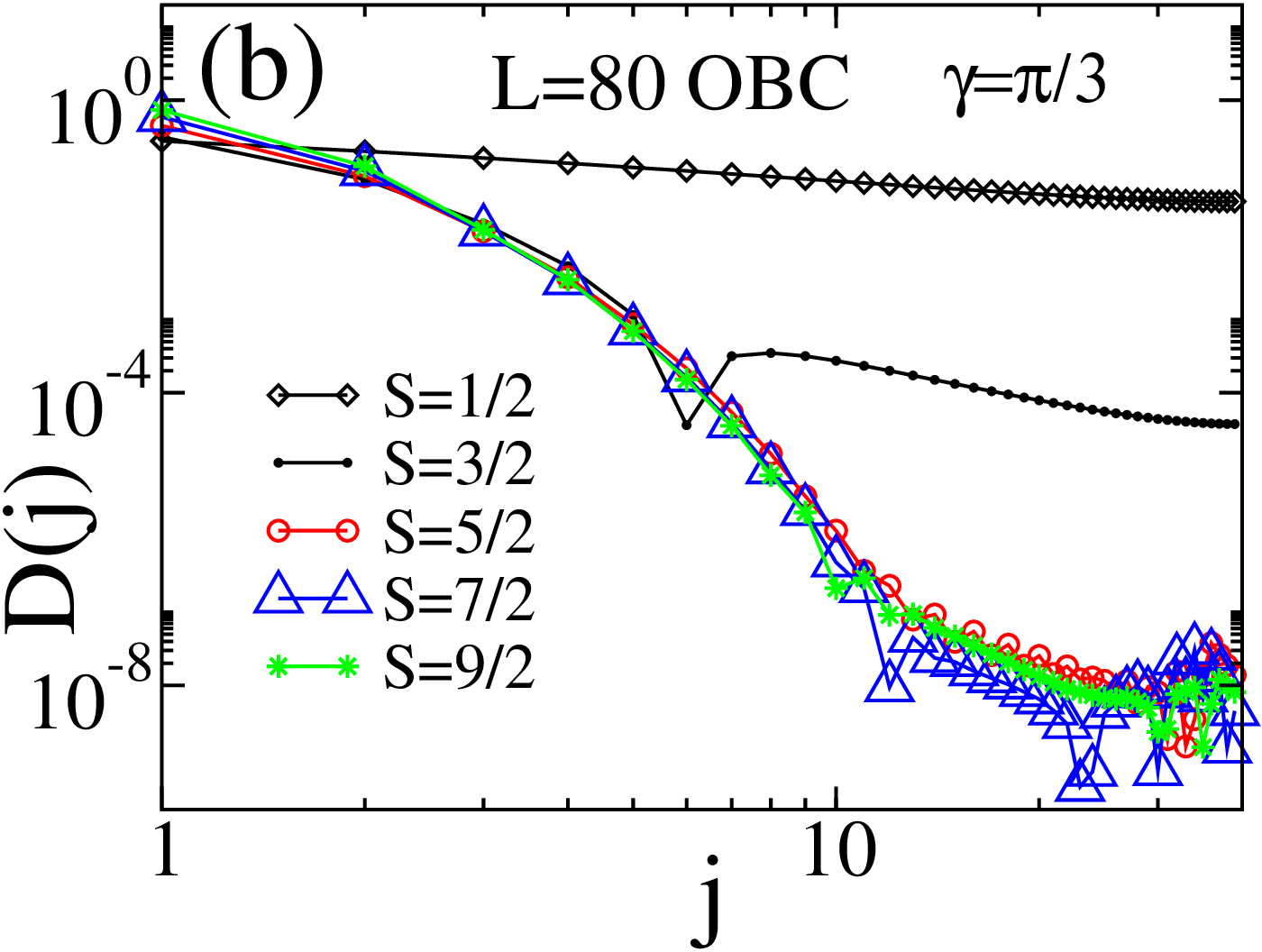}
\par\end{centering}

\caption{\label{fig3} (Color online). Some results for
the anisotropic spin-S Heisenberg with OBC,  $\gamma=\pi/3$ 
and some values of $S$ (see legend).
(a) The  entanglement entropy $S(L,l_{A})$ 
vs. $l_{A}$ for chains of size $L=160$.
 (b) The dimer parameter $D(j)$ vs. j for chains of size $L=80$. We present $D(j)$ only for
distances $j<40$.
 }

\end{figure}

Now, let us consider the OBC case.  In Fig. \ref{fig3}(a) we show
$S(L,l_{A})$ for the anisotropic spin-S Heisenberg with OBC, $L=160$, $\gamma=\pi/3$ and some values
of spins $S$. Note that differently of the spin-1/2 case with OBC, 
for spins $S>1/2$ the entanglement entropy $S(L,l_{A})$ does not 
present the strong odd-even oscillations.
Laflorencie {\it et al.} in Ref. \onlinecite{entropyaffleckosc} argue 
that the oscillating term  in $S(L,l_{A})$ is  universal, 
and is connected with the antiferromagnetic nature of the Hamiltonian
(however,  at the present moment, there is no analytical 
derivation of this statement). In particular, 
the alternating term in $S(L,l_{A})$  appears to be proportional 
to the alternating term in the energy density\cite{entropyaffleckosc}
 
\begin{equation}
h(j)=< (s_{j}^{x}s_{j+1}^{x}+s_{j}^{y}s_{j+1}^{y}+\Delta
s_{j}^{z}s_{j+1}^{z}) >. 
\label{eq:12}
\end{equation}

In order to understand the absence of the strong odd-even oscillations in
entanglement entropy $S(L,l_{A})$ for $S>1/2$, we study the behavior of 
the ``dimer parameter'' $D(j)=h(j+1)-h(j)$. 
It is expected that $D(j)$ decay with a power law.\cite{entropyaffleckosc} 
In Fig. \ref{fig3}(b), we present
the dimer parameter $D(j)$ for chains of size $L=80$. \cite{commdimer}
We roughly estimate the errors of $D(j)$ around $\sim 10^{-6}$, since we
find numerically  that $(E_0-\sum_j h(j))\sim 10^{-6}$.
As observed in this figure, the magnitude of the oscillations of $h(i)$ for $S>1/2$  
are very small. These results, indeed, corroborate that the alternating terms
in $S(L,l_{A})$ and $h(l_{A})$ are connected.
However, due to the fact that the values of $D(i)$ is so small for $S>1/2$,
we were not able to check that the alternate parts of $S(L,l_{A})$ and  $h(l_{A})$ decay with
the same exponent,  as expected by Laflorencie {\it et al.} \cite{entropyaffleckosc}

As in the PBC case, we also observed that the curves of the rescaled  entanglement entropies  $W(L,l_{A},S)$ 
of chains with $S\ge 3/2$ and  OBC    collapse onto a single universal scaling curve,
as we can see in Fig. \ref{fig2}(c).



Finally, let us present our estimates of $c^{PBC}$,  $x_{bulk}$ 
and $x_{s}$  for   several values of spins S. In Table III, 
a summary of our results is provided (acquired following the
procedure explained in the subsection II.A).
The finite-size estimates of $c^{PBC}$ and $x_{bulk}$ were obtained 
considering  systems of sizes $L=80$ or 
$L=96$ with PBC, while the extrapolated values of $c^{OBC}$ 
and $x_{s}$ were obtained considering
the model with OBC and sizes up to $L=180$. We also show in this
table the ratios of the dimensions $x_{s}/x_{bulk}.$ Based in our
benchmark results of the  spin-1/2 case, we believe the errors
of the quantities presented in table III are smaller than $10^{-3}$.

Note that it is expected that the critical behavior of the Heisenberg chains
with  half-integer
spins belongs to  the same class 
of the  Gaussian model.\citealp{chico1,spinschico,others}
In particular, numerical\citealp{chico1,karen,spinschico} and analytical\citealp{spinsaffleck,peschel2,others} techniques
show  that $c=1$  for the  Heisenberg chains
with  half-integer spins.  Moreover, 
it is expected that anomalous dimension of the  anisotropic Heisenberg chains
depend of the anisotropy and of the value of the spin $S$.\citealp{chico1,spinschico,peschel2,comm4a}

\begin{table}

\caption{Finite-size estimates of the conformal anomaly $c^{PBC}$ and the
anomalous dimension $x_{bulk}$ for the  anisotropic spin-S Heisenberg
chain with PBC and $\gamma=\pi/3$ obtained with $L=80$-96. The extrapolated
values of the conformal anomaly $c^{OBC}$ and the surface exponent
$x_{s}$ are also presented for the same model/coupling with OBC.}
\vspace*{0.2cm} 
\begin{tabular}{cccccc}
\toprule 
\midrule 
$S$ & $c^{PBC}$ & $c^{OBC}$ & $x_{bulk}$ & $x_{s}$ & $x_{s}$/$x_{bulk}$\tabularnewline
\midrule 
1/2 & 0.9997 & 1.004 & 0.3333 & 0.6665 & 1.999\tabularnewline
\vspace*{0.05cm}
3/2 & 0.9995  & 1.002 & 0.09918 & 0.1984 & 2.000\tabularnewline
\vspace*{0.05cm}
5/2 & 0.9993 & 0.999 & 0.0572 & 0.1143 & 1.998\tabularnewline
\vspace*{0.05cm}
7/2 & 0.9989 & 1.002 & 0.0403 & 0.0806 & 2.000\tabularnewline
\vspace*{0.05cm}
9/2 &  0.9995 & 1.001 & 0.0311 & 0.0623 & 2.003 \tabularnewline
\midrule 
\bottomrule
\end{tabular}
\end{table}

As observed in  Table III, our estimates of $c$  
are in agreement with the expected value of $c=1$.
Moreover, our results for  $S=3/2$ are in perfect agreement with the ones 
found by Alcaraz and Moreo.\citealp{spinschico} In particular, for $\gamma=\pi/3$
they found the following \emph{extrapolated} values: $c=1.08$, $x_{bulk}=0.098$
and $x_{s}=0.198$.
Our results are also consist with the Alcaraz-Moreo's conjecture
$x_{s}=2x_{bulk}$. This conjecture was proposed based in the exact
diagonalization calculations of small system sizes.
Besides that, Alcaraz and Moreo\citealp{spinschico} reported results 
only for spins up to $S=2$ for the antiferromagnetic  
region $0<\Delta \le 1$. \citealp{comm4}
Here, considering larger system sizes (giving better estimates) 
and larger values of spins we show that this conjecture holds.

\section{Conclusion}

In this article, we present a simply procedure to obtain  accurate estimates
of the conformal anomaly $c$, sound velocity $v_{s}$, anomalous
dimension $x_{bulk}$ and the surface exponent $x_{s}$ using the
conformal invariance relations and the density-matrix renormalization-group 
technique. In order to illustrate the procedure we use to get these
quantities, we investigate the anisotropic spin-S Heisenberg chains with 
periodic/open boundary conditions.

Our results for the model with spin S=1/2 were compared with the exact results.
For the spin-1/2 case, we found that the procedure that we used to estimate 
$c$, $v_{s}$, $x_{bulk}$ and $x_{s}$ gives  accurate estimates with errors
smaller than $10^{-3}$. We also present accurate results for the model with 
spins $S=3/2,$ 5/2, 7/2 and 9/2, and we confirm the Alcaraz-Moreo's 
conjecture $x_{s}=2x_{bulk}$. 

Our numerical results also support that   the entanglement
entropy of the spin-$S$ Heisenberg chains satisfies the relation
$S(L,l_{A},S)-S(L,l_{A},S-1)=1/(2S+1)$ for $S>3/2$ in the thermodynamic limit.
We also verified that the alternate terms of $S(L,l_{A})$ and $h(l_{A})$ seem
to actually be connected, as suggested by  Laflorencie {\it et al.}
\cite{entropyaffleckosc} However, we were unable to verify that both terms
decay with the same universal exponent. In this vein,
it is interesting to observe the universal oscillatory behavior of the R\'enyi
entropy  predicted by Calabrese and co-authors in
Ref. \onlinecite{entropyosc}. 
This study is in progress and the results will be present elsewhere.


%
%
%

\begin{acknowledgments}
This Research was supported by the Brazilian agencies FAPEMIG and CNPq. 
\end{acknowledgments}
\bibliographystyle{apsrev}

\begin{thebibliography}{28}
\bibitem{comm3} It is also possible to extract finite-size estimates of $v_s$ 
from Eq. 1 considering the system  with OBC. For this case it is necessary to
consider two systems with different sizes.
\bibitem{commdimer} We choose to present the results  for small system sizes,
since  it is expected the oscillations are  stronger for small system sizes.
\bibitem{comm4a} For the ferromagnetic region $-1< \Delta \le 0$ Alcaraz and
  Moreo  were able to find an analytical expression  for the dimensions of the
  Gaussian model (see Ref. \onlinecite{spinschico}).
\end{thebibliography}

\end{document}